# Phase Transition Pathway Sampling via Swarm Intelligence and Graph Theory

Li Zhu[†,*], R. E. Cohen[†,‡] and Timothy A. Strobel[†,*]

[†]Geophysical Laboratory, Carnegie Institution for Science, 5251 Broad Branch Road, NW, Washington, DC 20015, USA

[‡]Department of Earth and Environmental Sciences, Ludwig Maximilians Universität, Munich 80333, Germany

**ABSTRACT:** The prediction of reaction pathways for solid-solid transformations remains a key challenge. Here, we develop a pathway sampling method via swarm intelligence and graph theory, and demonstrate that our PALLAS method is an effective tool to help understand phase transformations in solid-state systems. The method is capable of finding low-energy transition pathways between two minima without having to specify any details of the transition mechanism *a priori*. We benchmarked our PALLAS method against known phase transitions in cadmium selenide (CdSe) and silicon (Si). PALLAS readily identifies previously-reported, low-energy phase transition pathways for the wurtzite to rock-salt transition in CdSe and reveals a novel lower-energy pathway that has not yet been observed. In addition, PALLAS provides detailed information that explains the complex phase transition sequence observed during the decompression of Si from high pressure. Given the efficiency to identify low-barrier-energy reaction pathways, the PALLAS methodology represents a promising tool for materials by design with valuable insights for novel synthesis.

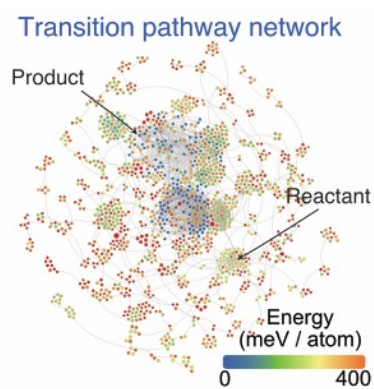

The discovery and implementation of novel materials plays an essential part in the development of modern technology. Distinct from the traditional, "Edisonian-type" discovery of new materials, computer-assisted methods, which reduce the number of trial-and-error experimental attempts, have been a powerful tool for the acceleration of materials design and discovery. The initial step in the theoretical design of a material is to evaluate "synthesizability" (i.e., energetic stability), which relies heavily on structural information. Crystal structure prediction (CSP) methods, which aim to find the ground-state configuration on the high-dimensional potential energy surface (PES), play a key role in materials design and discovery. Several structure prediction methods such as genetic algorithms[1], random sampling[2], Basin Hopping[3], Minima Hopping[4], and particle swarm optimization[5,6], have been developed to identify the most stable structures within the vast configurational space of the PES. By combining CSP methods with density functional theory (DFT), ground-state configurations may be determined for a given chemical system at a specific thermodynamic condition. In many systems, however, there are numerous configurations with energies only slightly above the ground state. Due to the vast number of these metastable states as compared with the unique ground state, any particular ground-state property is likely not the optimal one possible for that system. For example, the hardest form of carbon (diamond) is metastable at 1 atm. The metastable $TiO_2$ polymorph anatase is known to exhibit superior photocatalytic activity over its ground-state rutile counterpart.[7,8] Thus, the ability to control and direct the synthesis of metastable materials is necessary to achieve extraordinary control over properties for specific applications.

According to transition state theory, the thermodynamic properties of transition states (saddle points) between local minima can be used to quantify chemical reaction rates and to identify possible synthetic pathways.[9,10] While the limited knowledge of a collection of minima is important for materials discovery (c.f., CSP methods), a much more detailed understanding of the PES, including transition pathway details, is needed to further advance synthetic control and fundamental understanding of phase transitions.

A number of approaches have arisen to determine the reaction pathways that connect PES minima with each other. Several methods have been successfully applied on isolated systems, such as molecules or clusters.[11–19] For crystalline systems with periodic boundary conditions (PBC), one strategy is based on network representations, where phase transitions are treated as topological transformation networks.[20,21] This topological approach enables the prediction of new phases that are closely related to known phases and their associated transition pathways.[21] Another well-known pathway sampling scheme is based on the nudged elastic band (NEB) method,[22,23] which constructs a virtual "band" between the initial and final states and searches for the minimum energy transition pathway. In order to apply the NEB method, one must devise a suitable starting interpolation between the initial and final states, and identify each pair of corresponding atoms between the two states. Despite the various successful applications of the NEB method for examining transition mechanisms (e.g., simple diffusion), there are major limitations for the investigation of phase transitions involving complex crystal structures. The number of assignments of corresponding atoms between the initial and final states increases combinatorially with the number of atoms within the unit cell. Additionally, any choice of lattice points can serve as cell vectors for the same crystalline structure, which increases complexity of the transition mapping.

Recently, mode-following methods, such as the solid-state dimer method[24] or the crystal stochastic surface walking method[25], were developed to circumvent these issues. These mode-following methods establish paths from minima to first-order saddle points by following eigenmodes of the Hessian. Instead of optimizing an initial input pathway, mode-following methods naturally evolve configurations between saddle points and adjacent minima. However, mode-following methods are typically applied as a single-end approach, which means they cannot be directly used for extracting transition pathways between two specific states, and the energy landscape must be extensively explored when using this kind of approach.[25–28] Reactive processes theoretically can be obtained from molecular dynamics (MD), but standard MD simulations are limited by the time scales required for sampling rare events such as phase transitions. Transition path sampling (TPS) is a recent method to address the problem of rare events.[29,30] Nevertheless, TPS requires an initial path connecting the reactant and product, making it difficult to apply to systems for which no previously knowledge of PES is available.

To identify the best pathway (defined as the path with the global minimum barrier energy), one needs not only to determine the possible intermediate states that can connect between reactant and product, but one also needs to find the series of connections with the lowest barrier energy. Here, we present a new pathway sampling method based on swarm intelligence and graph theory (PALLAS) that is designed to efficiently find the transition states (TS) and the associated intermediate local minima (M) along the reaction pathway. The transition state is defined as a first-order saddle point of the energy function.[31] PALLAS integrates the solid-state dimer method (SSD) as the saddle point optimization engine[24], a robust "fingerprint" method[32] that can accurately quantify the dissimilarity between crystal structures, structure searching through multi-objective particle swarm optimization (MOPSO)[33], and graph theory[34] to automatically analyze a complex network of transition paths in order to find the "best" low-barrier pathways between two specific initial and final states for crystalline systems with PBC.

The PALLAS method, depicted in the flow chart (**Figure 1**), consists of five main parts: (1) generation of random velocities for both the reactant and product; (2) climbing processes to locate saddle points and near-by minima; (3) measurement of configurational distances between the different structures; (4) analysis of pathway networks using graph algorithms; (5) generation of new velocities by multi-objective particle swarm optimization (MOPSO) for minimizing the fingerprint distance and barrier energy.

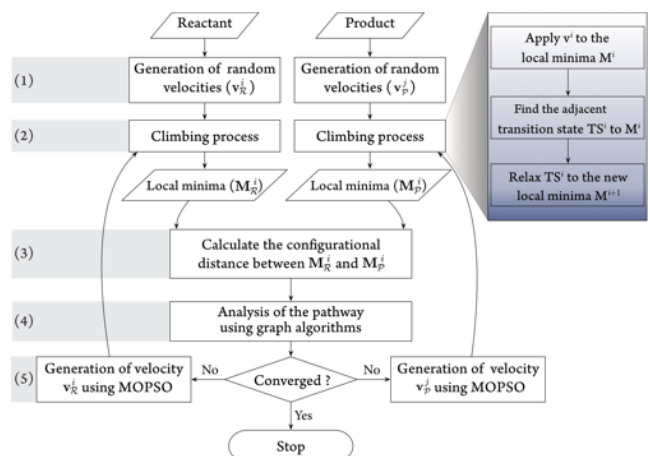

**Figure 1.** Flowchart of the PALLAS method.

The first step (1) of our approach is to generate random velocities for the reactant and product. In this work, the reactant and product are crystal structures, which are represented using two components: the basis vectors of the unit cell and the atomic coordinates within the cell. The velocities are used to change the cell vectors and atomic positions. Since the atomic positions and cell vectors have different units, we need to combine them into a unified velocity vector. Here we adopt the Jacobian metric, as introduced by Xiao et al.[24], to define a uniform velocity for a crystal structure,

$$\boldsymbol{v} = \{J\boldsymbol{v}_l, \boldsymbol{v}_a\}, \quad (1)$$

where $J$ is the Jacobian,[24] $\boldsymbol{v}_l$ is the lattice velocity describing the changes of the unit cell, and $\boldsymbol{v}_a$ is atomic velocity indicating the changes in the atomic coordinates. Note that $\boldsymbol{v}_l$ is a $3 \times 3$ matrix and $\boldsymbol{v}_a$ is a $N \times 3$ matrix, where N is the number of atoms in the unit cell, so $\boldsymbol{v}$ can be written as a $(N + 3) \times 3$ matrix.

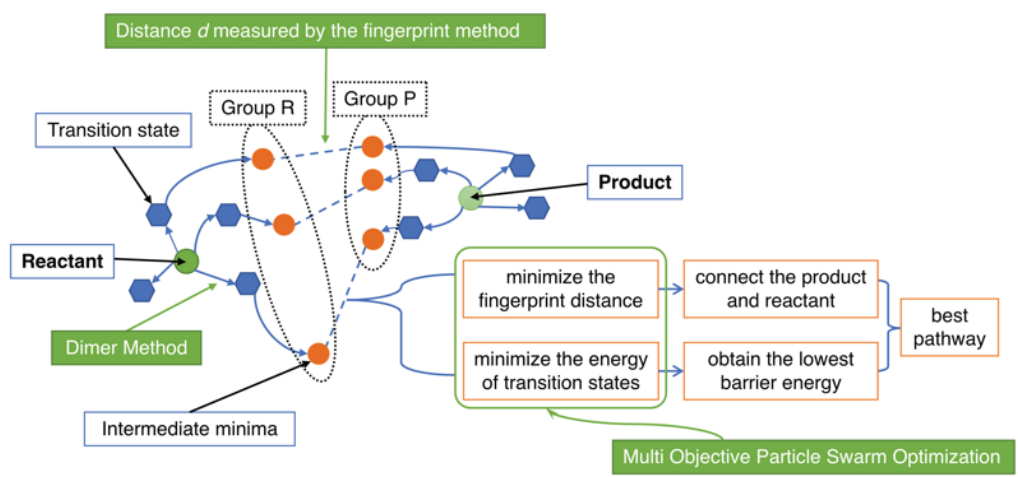

**Figure 2.** Schematic diagram of the PALLAS method. The blue hexagons depict the transition states obtained by using the SSD method. The orange circles indicate the intermediate minima from the geometry optimization of transition states. The MOPSO method optimizes the fingerprint distance d and the energies of the transition states. Once the fingerprint distance is minimized to "zero" (less than a threshold value, in practice), a transition pathway that connects the reactant and the product will be discovered.

In the second step (2), random velocities ($v$) are applied to the reactant and product. The random velocities are generated with a multivariate Gaussian distribution, which leads to two groups of configurations displaced from the starting minima. Starting with the two random reactant (**R**) and product (**P**) images, two groups of saddle points are generated along different directions by applying the SSD method, as shown in **Figure 2**. By following local optimization procedures, the two sets of saddle points naturally follow transition paths to close-by local minima. The minima from group **R** are connected to the reactant, and the minima from group **P** are connected to the product. We use an undirected network graph to represent the collection of local minima, transition states, and connections between them. Network graphs have been used previously to represent crystal structures and atomic bonds in crystals, and represent a powerful tool to investigate topological properties of crystalline systems.[20] Within the PALLAS method, the network graph is implemented to describe the relations between different structures rather than the topological relations of atoms within the same structure.[20] Within the network, each structure (minima and transition states) is considered as a node, while edges depict transition pathways between them.

Structures within the pathway network must be distinguished from each other. To estimate similarity/dissimilarity between different structures in the third step (3), we utilize a crystal fingerprint for measuring the configurational distances between the structures from group **R** and the structures from group **P**. A multitude of crystal fingerprint methods have been proposed for the analysis of the structural similarities, including Coulomb matrices,[35] Behler-Parrinello symmetry functions,[36] smooth overlap of atomic position kernels,[37] element resolved radical distribution functions[38] as implemented in the USPEX code,[1] and bond characterization matrices as in the CALYPSO code.[5,6] The crystal fingerprint technique used here is based on Gaussian overlap matrices,[32,39] which characterize the local environments of all atoms in a cell and can efficiently calculate configurational distances between crystalline structures that satisfy the mathematical properties of a metric[32]. This particular fingerprint method was demonstrated to be robust over a wide range of structure types and clusters.[32] Here, configurations are considered to be identical if their fingerprint distance is < $10^{-3}$.

Extracting the lowest barrier paths with the least number of intermediate transition states from the pathway network is related to the minimax path problem in graph theories. This path problem is a subject of combinatorial optimization, and exhaustive search is not tractable. In the fourth step (4), we apply Kruskal's algorithm[40] and breadth-first search[41,42] (BFS) to analyze the pathway network and estimate the barrier energy $E_b(X_i)$ between the local minima $X_i$ and reactant/product. First, we utilize Kruskal's algorithm to find paths that connect both minima with the lowest possible energy barrier $E_m$. Next, we truncate the pathway network by removing all transition states with energies higher than $E_m$. We then use BFS to pass through the truncated network and find the optimal transition pathways with the lowest barrier during the searching process.

In the fifth step (5), we utilize the advantages of MOPSO[33] to discover the optimal pathways automatically. MOPSO is a robust stochastic optimization technique based on swarm intelligence. It utilizes the hypothesis of social interaction for problem-solving. Within the MOPSO scheme, a configuration in the search space is treated as a particle. For particle $i$, we combine the fingerprint distances and structural energies to define the fitness function as

$$F_{fitness}^i = \frac{w_1 \min_j \{D(X_{i,j})\} + w_2 E_b(X_i)}{w_1 + w_2}$$

(2)

where $X_i$ is the configuration of structure $i$, $D(X_{i,j})$ is fingerprint distance from structure $i$ to structure $j$ of the opposite group, $E_b(X_i)$ is the barrier energy of the pathway from structure $i$ to the reactant/product, and $w_1$ and $w_2$ are the weights for the two parts.

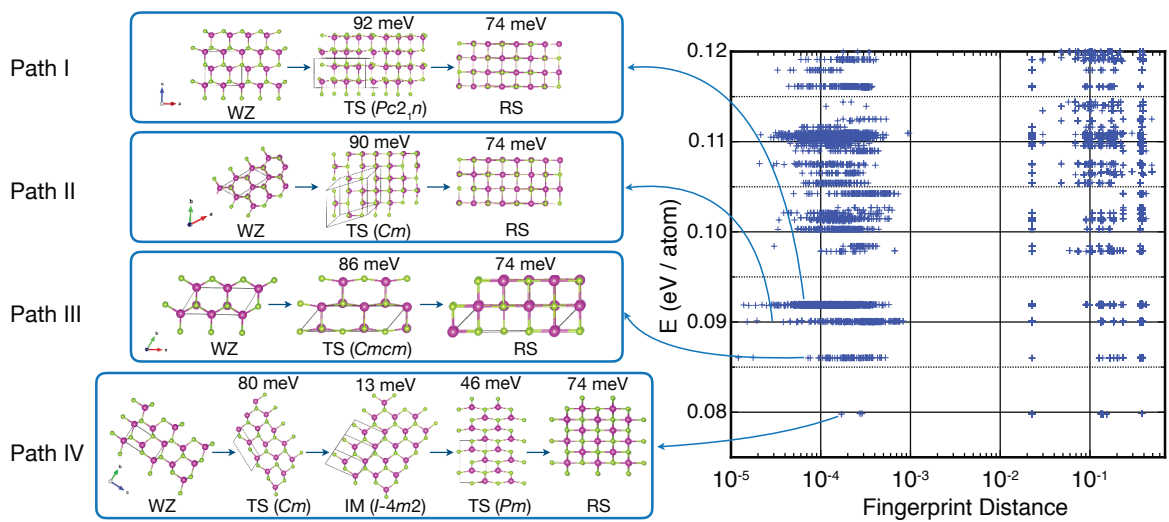

**Figure 3.** Low-energy WZ-RS transition pathways obtained for the CdSe system. Energy values are in meV/atom with respect to the WZ structure. The left panel shows four transition pathways discovered using the PALLAS method. In the right panel, the horizontal axis shows the fingerprint distance (d) between two closest intermediate minima from the reactant and the product side respectively, and the vertical axis indicates the barrier energies. Here, fingerprint distances < 10$^{-3}$ indicate that the two closest intermediate minima are identical and thus the RS and WZ phases are connected through a pathway of intermediate structures. Multiple points with fingerprint distances < 10$^{-3}$ at one energy indicate the sampling of identical (within numerical tolerance) pathways. The fingerprint distance value is not precisely zero due to numerical tolerances of geometry optimization for atomic positions and cell coordinates.

The fitness function is then minimized by the MOPSO algorithm, which is designed to perform multidimensional optimization. The new velocity of each particle ($i$) at the PSO iteration step $t$+1 is determined by:

$$\boldsymbol{v}_i^{t+1} = \omega \boldsymbol{v}^t + c_1 r_1 (\boldsymbol{pbest}_i^t - \boldsymbol{x}_i^t) + c_2 r_2 (\boldsymbol{gbest}_i^t - \boldsymbol{x}_i^t) \quad (3)$$

where $\omega$ denotes the inertia weight, $c_1$ and $c_2$ are learning factors, $r_1$ and $r_2$ are two separately generated random numbers in the range [0, 1], and $\boldsymbol{x}_i^t$ is the configuration of particle $i$ at step $t$. Previous studies suggested that $c_1 = c_2 = 2$ and $0.4 \leq \omega \leq 0.9$ gave the best overall performance for PSO[5,6,43]. At any particular step, the particle in the swarm with the best fitness value will be selected as the global best particle (***gbest***), while each individual particle maintains a personal best position (***pbest***$_i$) achieved to date.

Instead of updating the position of particles by the original PSO formula $\boldsymbol{x}_i^{t+1} = \boldsymbol{x}_i^t + \boldsymbol{v}_i^{t+1}$, here the new velocities generated are applied to the minima through SSD calculations, leading to new sets of transition states and subsequent adjacent minima. The movement of each particle in the search space is dynamically influenced by its individual past experience (***pbest***) and global best experience attained by the entire swarm (***gbest***) (Eq. 3). Thus, the evolution in velocity makes the particles move towards the global best solution, which is the transition pathway with the lowest barrier between the reactant and the product.

The PALLAS simulation is stopped when the halting criterion is reached. Similar to the problem of structure prediction, the determination of the transition pathway with the global minimum barrier energy can be classified as an NP-hard (non-deterministic polynomial-time hard) problem, which means there are an *infinite* number of transition pathways; it is impossible to develop an algorithm to determine *all* pathways. The development/utilization of accurate reactive force fields (e.g., machine learning force fields[36,44]) combined with growing computing power will tend towards improved search efficiency, however there is no absolute criterion that can be applied for halting the PALLAS method. In practice, default halting occurs after the simulation does not find a new lower-energy-barrier transition pathway after 10 consecutive generations.

The PALLAS method can currently interface with several external packages to perform energy and force calculations, e.g., VASP[45] and LAMMPS[46]. PALLAS is very amenable to parallel implementation as calculations for each particle are virtually independent and little communication is required between walkers. By parallelizing the PALLAS method, pathway searches run very efficiently on modern parallel computers. We now describe two applications of PALLAS to illustrate the utility of the method.

In the first application, we studied transition pathways for crystalline CdSe. In recent years, many computational approaches have been utilized to explore phase transitions from the wurtzite (WZ) to rock-salt (RS) structures in binary semiconductors[22,24], and comparison with the PALLAS method serves as a useful benchmark. We examined pathways for the WZ → RS phase transition in CdSe using the same empirical potential[47] within LAMMPS[46] for direct comparison with previous studies[22,24].

The CdSe simulation results are summarized in **Figure 3**, in which a selected set of transition pathways found by PALLAS are presented. Using the PALLAS method, we successfully reproduced the two low-energy transition pathways obtained previously.[22,24] These previously-determined pathways, with barrier energies of 92 meV/atom (path I) and 86 meV/atom (path III), respectively, were discovered using the SSD method combined with a kinetic Monte

Carlo algorithm[22,24]. Interestingly, a new mechanism with a lower barrier was found for the WZ to RS transition (path IV), which has not been reported before. This newly discovered path IV involves two transition states and one intermediate minimum. First, a displacement along the $b$-axis of the WZ structure results in a new transition state with the space group $Cm$. Next, a stacking rearrangement within the $Cm$ structure leads to an intermediate minimum, which is a body-centered tetragonal structure (bct, $I$-$4m2$). Ultimately, the RS structure is formed through a contraction of the bct structure via another transition state with space group $Pm$. This newly discovered low-energy transition pathway demonstrates that the PALLAS method is capable of exploring the energy landscape very efficiently.

In addition, we found new pathways with slightly higher barrier energies by using our PALLAS method (e.g., path II with the barrier energy of 90 meV). The underlying transition mechanism of both path II and path III can be identified as a shearing of (110) layers in the WZ phase, which is consistent with previous MD simulations.[48–50] The method may sample the same transition pathways during the calculation, as indicated by the blue points with the same energies shown in **Figure 3**. Compared to the pathways with higher barriers, the 80 meV / atom pathway is sampled much less frequently, which means it is more difficult obtain on the PES. The discovery of this low-barrier pathway shows that the PALLAS method has the ability to sidestep getting stuck in deep funnels.

For the second application, we explored the transition pathways of silicon. In this case, density functional theory (DFT) was used to evaluate the energies and forces. The DFT calculations were performed within the Perdew-Burke-Ernzerhof exchange-correlation[51] as implemented in the VASP code[45]. Projector-augmented wave potentials (PAW)[52] were used with valence electrons as $3s^2 3p^2$. The simulation cell was chosen to contain up to 16 atoms using a dense Monkhorst–Pack $k$ mesh[53] to sample the Brillouin zone.

As the most important semiconductor material used widely for electronic devices, silicon has become the cornerstone of our technology-based society. While diamond cubic is the most stable structure at ambient pressure, silicon exhibits a rich and well-documented phase diagram under high pressures[54]. Numerous allotropes have been discovered under high pressures either by theoretical or experimental studies[55–58], and several of them can be stabilized at ambient conditions[59–61, 62]. In addition, multiple silicon phases are known to form through complex kinetically-controlled processes[63].

Despite a great number of previous theoretical studies on the energetic stabilities of different Si structures, there are very few studies regarding the atomistic mechanisms and transition pathways for these complicated phase transitions. One fascinating example is the irreversible metastable back-transformation associated with decompression from the high-pressure β-Sn phase. Instead of returning to the most energetically stable diamond structure ($d$-Si), the β-Sn structure transforms into metastable BC8/R8 structures upon decompression.[59–61,64–66] Wang et al.[67] examined the transition mechanism for this irreversible structural transformation using the NEB method[22]. The pathway from β-Sn to BC8 was found to have a much lower barrier energy (e.g., ~150 meV / atom at 8 GPa) than the pathways from β-Sn to other Si phases, which helps explain the unusual avoidance of the thermodynamically stable phase.

We also searched for transition pathways for the β-Sn → BC8 in Si at 8 GPa using our PALLAS method. We discovered a transition pathway with a much lower barrier of 96 meV/atom at 8 GPa as shown in **Figure 4**. Interestingly, this pathway also contains the R8 phase, which serves as an intermediate minimum between the BC8 and β-Sn structures. This observation explains why the β-Sn phase transforms first to the R8 structure before converting to the BC8 structure in decompression experiments[60,61,66,68–71]. The relative energies of these phases shift with pressure and can favor the formation of either R8/BC8 depending on the depressurization conditions. That is, at 8 GPa the R8 structure is the global minimum along this pathway, and BC8 becomes the lower-energy structure at lower pressure.

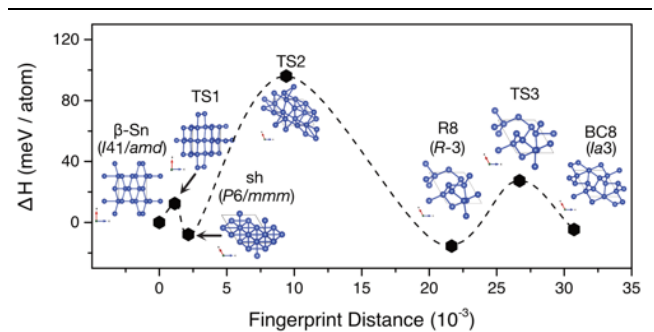

**Figure 4.** Phase transition pathway for β-Sn → BC8 Si at 8 GPa.

On heating the BC8 structure at ambient pressure, it transforms to yet another metastable phase, hexagonal diamond silicon ($hd$-Si), rather than converting back to the most stable $d$-Si structure[59]. To our best knowledge, no theoretical study has been performed to investigate this phase transformation process. By using our PALLAS method in combination of first principles calculation, we searched for transition pathways between BC8 → $hd$-Si and BC8 → $d$-Si at ambient pressure. The calculation results are summarized in **Figure 5**, which includes all of the local minima and transition states from the sampling. Here we use a network representation to organize the complex transition pathway network, which shows the connected pathways between different structures of silicon at ambient pressure. The network layout was calculated using the ForceAtlas 2 algorithm[72] implemented in the Gephi package[73].

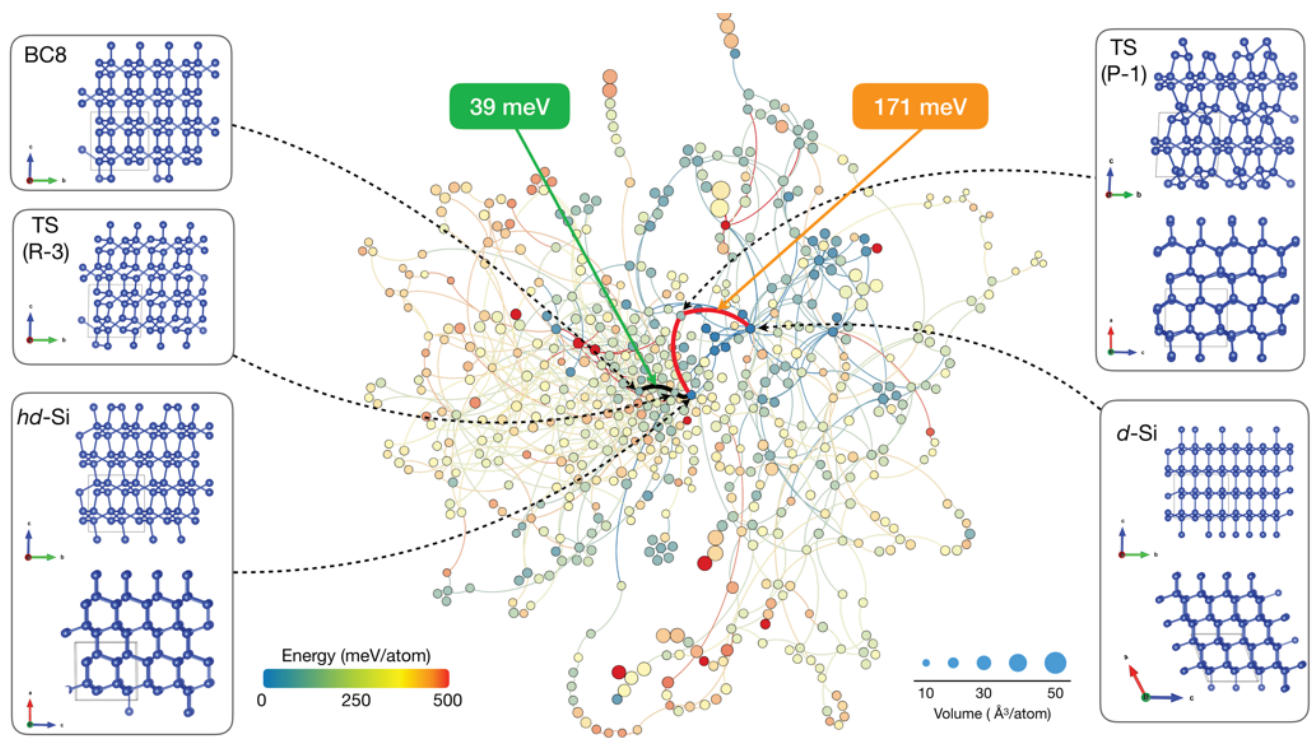

**Figure 5.** Transition pathway network for Silicon at 1 atm predicted by using the PALLAS method. Each circle represents a unique structure (including local minima and saddle points). The size and color represent volume and energy, respectively. The thin lines show all the different pathways between the different structures. The thick lines shows the "best" (lowest energy) pathway from BC8 → d-Si, and hd-Si is an intermediate minimum along the transition pathway. For clarity, the overall path is broken into two colors (black and red) to indicate the portions for BC8 → hd-Si and hd-Si → d-Si, respectively. The reactant, product and transitions state structures are depicted on the sides.

Intriguingly, we find that hd-Si is located on the lowest-energy transition pathway from BC8 → d-Si as an intermediate minimum. For BC8 → hd-Si, the barrier energy is estimated as ~39 meV / atom, and the phase transition involves local bond rotations. This low-energy kinetic barrier suggests that local bond rotations can easily take place at moderate temperature. The calculated energy barrier ($k_B$*450 K) is consistent with the experimental transition temperature above 200 °C (473 K).[59] However, the conversion from hd-Si to d-Si is associated with bond twisting and breaking, resulting in a high barrier energy of ~171 meV /atom. Our results indicate that the conversion from BC8 to d-Si is inhibited by the high kinetic barrier of hd-Si → d-Si, which is consistent with the experimental findings.[59] The transition from hd-Si to d-Si occurs upon heating above 600 °C.[74]

In summary, we have developed the PALLAS method that aims to find transition pathways for crystalline systems automatically. The PALLAS method couples several major modules to efficiently explore the PES for constructing fully-connected transition pathway networks that are guided by configurational distance, and optimized by MOPSO and graph algorithms. By following SSD and local optimization procedures, the search will naturally follow transition paths on the PES. The PALLAS method is capable of discovering transition pathways between a given reactant and product without the predefinition of a reaction coordinate, or the use of prohibitive MD simulations for rare event sampling. The efficiency of our new method is confirmed by new results that were found in CdSe and Si phase transitions. The capabilities of the PALLAS method are promising as a major tool for materials by design and can also be used for generating large training sets (including both local minima and transition states) for machine learning techniques to predict materials properties. In addition, the transition pathway network provides many metastable structures with kinetic information, which is useful for screening whether new phases can be synthesized[75].

## AUTHOR INFORMATION

**Corresponding Author**

* E-mail: z@zhuli.name (L.Z.)
* E-mail: tstrobel@ciw.edu (T.A.S.)

## ACKNOWLEDGMENT

This work was supported by DARPA under Grant No. W31P4Q1310005. Computations were carried out at the supercomputer Copper of DoD HPCMP Open Research Systems under project No. ACOMM35963RC1 and the Memex cluster of Carnegie Institution for Science. REC also acknowledges support from the European Research Council Advanced Grant ToMCaT.

## REFERENCES

(1) Glass, C. W.; Oganov, A. R.; Hansen, N. USPEX—Evolutionary Crystal Structure Prediction. Comput. Phys. Commun. 2006, 175, 713–720.
(2) Pickard, C. J.; Needs, R. J. Ab Initio Random Structure Searching. J. Phys. Condens. Matter 2011, 23, 53201.
(3) Wales, D. J.; Miller, M. A.; Walsh, T. R. Archetypal Energy Landscapes. Nature 1998, 394, 758–760.
(4) Amsler, M.; Goedecker, S. Crystal Structure Prediction Using the Minima Hopping Method. J. Chem. Phys. 2010, 133, 224104.
(5) Wang, Y.; Lv, J.; Zhu, L.; Ma, Y. Crystal Structure Prediction via Particle-Swarm Optimization. Phys. Rev. B 2010, 82, 94116.
(6) Wang, Y.; Lv, J.; Zhu, L.; Ma, Y. CALYPSO: A Method for Crystal Structure Prediction. Comput. Phys. Commun. 2012, 183, 2063–2070.
(7) O'Regan, B.; Grätzel, M. A Low-Cost, High-Efficiency Solar Cell Based on Dye-Sensitized Colloidal $TiO_2$ Films. Nature 1991, 353, 737–740.
(8) Smith, S. J.; Stevens, R.; Liu, S.; Li, G.; Navrotsky, A.; Boerio-Goates, J.; Woodfield, B. F. Heat Capacities and Thermodynamic Functions of $TiO_2$ Anatase and Rutile: Analysis of Phase Stability. Am. Mineral. 2009, 94, 236–243.
(9) Evans, M. G.; Polanyi, M. Some Applications of the Transition State Method to the Calculation of Reaction Velocities, Especially in Solution. Trans. Faraday Soc. 1935, 31, 875.
(10) Eyring, H. The Activated Complex in Chemical Reactions. J. Chem. Phys. 1935, 3, 107–115.
(11) Henkelman, G.; Jónsson, H. A Dimer Method for Finding Saddle Points on High Dimensional Potential Surfaces Using Only First Derivatives. J. Chem. Phys. 1999, 111, 7010–7022.
(12) Heyden, A.; Bell, A. T.; Keil, F. J. Efficient Methods for Finding Transition States in Chemical Reactions: Comparison of Improved Dimer Method and Partitioned Rational Function Optimization Method. J. Chem. Phys. 2005, 123, 224101.
(13) Schaefer, B.; Mohr, S.; Amsler, M.; Goedecker, S. Minima Hopping Guided Path Search: An Efficient Method for Finding Complex Chemical Reaction Pathways. J. Chem. Phys. 2014, 140, 214102.
(14) Habershon, S. Sampling Reactive Pathways with Random Walks in Chemical Space: Applications to Molecular Dissociation and Catalysis. J. Chem. Phys. 2015, 143, 94106.
(15) Kim, Y.; Choi, S.; Kim, W. Y. Efficient Basin-Hopping Sampling of Reaction Intermediates through Molecular Fragmentation and Graph Theory. J. Chem. Theory Comput. 2014, 10, 2419–2426.
(16) Devaurs, D.; Vaisset, M.; Siméon, T.; Cortés, J. A Multi-Tree Approach to Compute Transition Paths on Energy Landscapes. 2013, 8–13.
(17) van Erp, T. S.; Moroni, D.; Bolhuis, P. G. A Novel Path Sampling Method for the Calculation of Rate Constants. J. Chem. Phys. 2003, 118, 7762–7774.
(18) Shang, C.; Liu, Z. Stochastic Surface Walking Method for Structure Prediction and Pathway Searching. J. Chem. Theory Comput. 2013, 9, 1838–1845.
(19) Zhang, X. J.; Liu, Z. P. Reaction Sampling and Reactivity Prediction Using the Stochastic Surface Walking Method. Phys. Chem. Chem. Phys. 2015, 17, 2757–2769.
(20) Blatov, V. A. Topological Relations between Three-Dimensional Periodic Nets. I. Uninodal Nets. Acta Crystallogr. Sect. A Found. Crystallogr. 2007, 63, 329–343.
(21) Blatov, V. A.; Golov, A. A.; Yang, C.; Zeng, Q.; Kabanov, A. A. Network Topological Model of Reconstructive Solid-State Transformations. Sci. Rep. 2019, 9, 6007.
(22) Sheppard, D.; Xiao, P.; Chemelewski, W.; Johnson, D. D.; Henkelman, G. A Generalized Solid-State Nudged Elastic Band Method. J. Chem. Phys. 2012, 136, 74103.
(23) Qian, G.-R.; Dong, X.; Zhou, X.-F.; Tian, Y.; Oganov, A. R.; Wang, H.-T. Variable Cell Nudged Elastic Band Method for Studying Solid–Solid Structural Phase Transitions. Comput. Phys. Commun. 2013, 184, 2111–2118.
(24) Xiao, P.; Sheppard, D.; Rogal, J.; Henkelman, G. Solid-State Dimer Method for Calculating Solid-Solid Phase Transitions. J. Chem. Phys. 2014, 140, 174104–174106.
(25) Shang, C.; Zhang, X.-J.; Liu, Z.-P. Stochastic Surface Walking Method for Crystal Structure and Phase Transition Pathway Prediction. Phys. Chem. Chem. Phys. 2014, 16, 17845–17856.
(26) Chill, S. T.; Welborn, M.; Terrell, R.; Zhang, L.; Berthet, J.-C.; Pedersen, A.; Jónsson, H.; Henkelman, G. EON: Software for Long Time Simulations of Atomic Scale Systems. Model. Simul. Mater. Sci. Eng. 2014, 22, 55002–55016.
(27) Chill, S. T.; Henkelman, G. Molecular Dynamics Saddle Search Adaptive Kinetic Monte Carlo. J. Chem. Phys. 2014, 140, 214110.
(28) Xie, Y.-P.; Zhang, X.-J.; Liu, Z.-P. Graphite to Diamond: Origin for Kinetics Selectivity. J. Am. Chem. Soc. 2017, 139, 2545–2548.
(29) Dellago, C.; Bolhuis, P. G.; Chandler, D. Efficient Transition Path Sampling: Application to Lennard-Jones Cluster Rearrangements. J. Chem. Phys. 1998, 108, 9236–9245.
(30) Bolhuis, P. G.; Chandler, D.; Dellago, C.; Geissler, P. L. Transition Path Sampling: Throwing Ropes Over Rough Mountain Passes, in the Dark. Annu. Rev. Phys. Chem. 2002, 53, 291–318.
(31) Wales, D. Energy Landscapes: Applications to Clusters, Biomolecules and Glasses; Cambridge University Press: Cambridge, U.K., 2003.
(32) Zhu, L.; Amsler, M.; Fuhrer, T.; Schaefer, B.; Faraji, S.; Rostami, S.; Ghasemi, S. A.; Sadeghi, A.; Grauzinyte, M.; Wolverton, C.; et al. A Fingerprint Based Metric for Measuring Similarities of Crystalline Structures. J. Chem. Phys. 2016, 144, 34203.
(33) Coello, C. A. C.; Pulido, G. T.; Lechuga, M. S. Handling Multiple Objectives with Particle Swarm Optimization. IEEE Trans. Evol. Comput. 2004, 8, 256–279.
(34) Sedgewick, R. Algorithms in C, Part 5: Graph Algorithms; Addison-Wesley: Boston, MA, 2001.
(35) Rupp, M.; Tkatchenko, A.; Müller, K.-R.; von Lilienfeld, O. A. Fast and Accurate Modeling of Molecular Atomization Energies with Machine Learning. Phys. Rev. Lett. 2012, 108, 058301.
(36) Behler, J.; Parrinello, M. Generalized Neural-Network Representation of High-Dimensional Potential-Energy Surfaces. Phys. Rev. Lett. 2007, 98, 146401.
(37) Bartók, A. P.; Kondor, R.; Csányi, G. On Representing Chemical Environments. Phys. Rev. B 2013, 87, 184115.
(38) Oganov, A. R.; Valle, M. How to Quantify Energy Landscapes of Solids. J. Chem. Phys. 2009, 130, 104504–104509.
(39) Sadeghi, A.; Ghasemi, S. A.; Schaefer, B.; Mohr, S.; Lill, M. A.; Goedecker, S. Metrics for Measuring Distances in Configuration Spaces. J. Chem. Phys. 2013, 139, 184118.
(40) Kruskal, J. B. On the Shortest Spanning Subtree of a Graph and the Traveling Salesman Problem. Proc. Am. Math. Soc. 1956, 7, 48.
(41) Moore, E. F. The Shortest Path through a Maze; Harvard University Press: Cambridge, MA, 1959.
(42) Skiena, S. S. The Algorithm Design Manual; Springer: London, U.K., 2008.
(43) Eberhart, R. C.; Shi, Y. Comparing Inertia Weights and Constriction Factors in Particle Swarm Optimization; IEEE: Piscataway, NJ, 2000.
(44) Bartók, A. P.; Payne, M. C.; Kondor, R.; Csányi, G. Gaussian Approximation Potentials: The Accuracy of Quantum Mechanics, without the Electrons. Phys. Rev. Lett. 2010, 104, 136403.
(45) Kresse, G.; Furthmüller, J. Efficient Iterative Schemes for Ab Initiototal-Energy Calculations Using a Plane-Wave Basis Set. Phys. Rev. B 1996, 54, 11169–11186.
(46) Plimpton, S. Fast Parallel Algorithms for Short – Range Molecular Dynamics. J. Comput. Phys. 1995, 117, 1–19.
(47) Rabani, E. An Interatomic Pair Potential for Cadmium Selenide. J. Chem. Phys. 2002, 116, 258.

(48) Zahn, D.; Grin, Y.; Leoni, S. Mechanism of the Pressure-Induced Wurtzite to Rocksalt Transition of CdSe. Phys. Rev. B 2005, 72, 064110.
(49) Shimojo, F.; Kodiyalam, S.; Ebbsjö, I.; Kalia, R. K.; Nakano, A.; Vashishta, P. Atomistic Mechanisms for Wurtzite-to-Rocksalt Structural Transformation in Cadmium Selenide under Pressure. Phys. Rev. B 2004, 70, 184111.
(50) Grünwald, M.; Dellago, C.; Geissler, P. L. An Efficient Transition Path Sampling Algorithm for Nanoparticles under Pressure. J. Chem. Phys. 2007, 127, 154718.
(51) Perdew, J. P.; Burke, K.; Ernzerhof, M. Generalized Gradient Approximation Made Simple. Phys. Rev. Lett. 1996, 77, 3865–3868.
(52) Blöchl, P. E. Projector Augmented-Wave Method. Phys. Rev. B 1994, 50, 17953–17979.
(53) Monkhorst, H. J.; Pack, J. D. Special Points for Brillouin-Zone Integrations. Phys. Rev. B 1976, 13, 5188–5192.
(54) Mujica, A.; Rubio, A.; Muñoz, A.; Needs, R. J. High-Pressure Phases of Group-IV, III V, and II VI Compounds. Rev. Mod. Phys. 2003, 75, 863–912.
(55) Botti, S.; Flores-Livas, J. A.; Amsler, M.; Goedecker, S.; Marques, M. A. L. Low-Energy Silicon Allotropes with Strong Absorption in the Visible for Photovoltaic Applications. Phys. Rev. B 2012, 86, 121204.
(56) Xiang, H.; Huang, B.; Kan, E.; Wei, S.-H.; Gong, X. Towards Direct-Gap Silicon Phases by the Inverse Band Structure Design Approach. Phys. Rev. Lett. 2013, 110, 118702.
(57) Wang, Q.; Xu, B.; Sun, J.; Liu, H.; Zhao, Z.; Yu, D.; Fan, C.; He, J. Direct Band Gap Silicon Allotropes. J. Am. Chem. Soc. 2014, 136, 9826–9829.
(58) Malone, B. D.; Cohen, M. L. Prediction of a Metastable Phase of Silicon in the Ibam Structure. Phys. Rev. B 2012, 85, 024116.
(59) Wentorf, R. H.; Kasper, J. S. Two New Forms of Silicon. Science 1963, 139, 338–339.
(60) Crain, J.; Ackland, G. J.; Maclean, J. R.; Piltz, R. O.; Hatton, P. D.; Pawley, G. S. Reversible Pressure-Induced Structural Transitions between Metastable Phases of Silicon. Phys. Rev. B 1994, 50, 13043–13046.
(61) Piltz, R. O.; MacLean, J. R.; Clark, S. J.; Ackland, G. J.; Hatton, P. D.; Crain, J. Structure and Properties of Silicon XII: A Complex Tetrahedrally Bonded Phase. Phys. Rev. B 1995, 52, 4072–4085.
(62) Kim, D. Y.; Stefanoski, S.; Kurakevych, O. O.; Strobel, T. A. Synthesis of an Open-Framework Allotrope of Silicon. Nat. Mater. 2014, 14, 169–173.
(63) Haberl, B.; Strobel, T. A.; Bradby, J. E. Pathways to Exotic Metastable Silicon Allotropes. Appl. Phys. Rev. 2016, 3, 040808.
(64) Kurakevych, O. O.; Le Godec, Y.; Crichton, W. A.; Guignard, J.; Strobel, T. A.; Zhang, H.; Liu, H.; Coelho Diogo, C.; Polian, A.; Menguy, N.; et al. Synthesis of Bulk BC8 Silicon Allotrope by Direct Transformation and Reduced-Pressure Chemical Pathways. Inorg. Chem. 2016, 55, 8943–8950.
(65) Zhang, H.; Liu, H.; Wei, K.; Kurakevych, O. O.; Le Godec, Y.; Liu, Z.; Martin, J.; Guerrette, M.; Nolas, G. S.; Strobel, T. A. BC8 Silicon (Si-III) Is a Narrow-Gap Semiconductor. Phys. Rev. Lett. 2017, 118, 146601.
(66) Wong, S.; Haberl, B.; Johnson, B. C.; Mujica, A.; Guthrie, M.; McCallum, J. C.; Williams, J. S.; Bradby, J. E. Formation of an R8-Dominant Si Material. Phys. Rev. Lett. 2019, 122, 105701.
(67) Wang, J.-T.; Chen, C.; Mizuseki, H.; Kawazoe, Y. Kinetic Origin of Divergent Decompression Pathways in Silicon and Germanium. Phys. Rev. Lett. 2013, 110, 165503.
(68) Weppelmann, E. R.; Field, J. S.; Swain, M. V. Observation, Analysis, and Simulation of the Hysteresis of Silicon Using Ultra-Micro-Indentation with Spherical Indenters. J. Mater. Res. 1993, 8, 830–840.
(69) Zarudi, I.; Zhang, L. C.; Cheong, W. C. D.; Yu, T. X. The Difference of Phase Distributions in Silicon after Indentation with Berkovich and Spherical Indenters. Acta Mater. 2005, 53, 4795–4800.
(70) Domnich, V.; Gogotsi, Y.; Dub, S. Effect of Phase Transformations on the Shape of the Unloading Curve in the Nanoindentation of Silicon. Appl. Phys. Lett. 2000, 76, 2214–2216.
(71) Kailer, A.; Gogotsi, Y. G.; Nickel, K. G. Phase Transformations of Silicon Caused by Contact Loading. J. Appl. Phys. 1997, 81, 3057–3063.
(72) Jacomy, M.; Venturini, T.; Heymann, S.; Bastian, M. ForceAtlas2, a Continuous Graph Layout Algorithm for Handy Network Visualization Designed for the Gephi Software. PLoS One 2014, 9, e98679.
(73) Bastian, M.; Heymann, S.; Jacomy, M. Gephi: An Open Source Software for Exploring and Manipulating Networks. Icwsm 2009, 8, 361–362.
(74) Kobliska, R. J.; Solin, S. A. Raman Spectrum of Wurtzite Silicon. Phys. Rev. B 1973, 8, 3799–3802.
(75) Jansen, M. A Concept for Synthesis Planning in Solid-State Chemistry. Angew. Chem. Int. Ed. 2002, 41, 3746–3766.